\documentclass[twocolumn,showpacs,superscriptaddress,longbibliography,nofootinbib,prd]{revtex4-2}
\usepackage[utf8]{inputenc}
\usepackage{calrsfs}
\usepackage{graphicx}
\usepackage{textcomp}
\usepackage{soul}
\usepackage{amsmath,amssymb}
\usepackage{color}
\usepackage{braket}
\usepackage{xcolor}
\usepackage{float}
\usepackage{physics}
\usepackage{soul}
\usepackage{graphicx}
\usepackage{caption}
\usepackage{amsmath}
\usepackage{amssymb}
\usepackage{bbold}
\usepackage{makecell}
\usepackage{tikz,pgfplots}
\usepackage{pgfplots}
\pgfplotsset{compat=1.18}
\usepgfplotslibrary{fillbetween}
\newcommand{\Dmax}{D_{\max}}
\newcommand{\calC}{\mathcal C}
\newcommand{\calE}{\mathcal E}

\usetikzlibrary{decorations.pathmorphing, arrows.meta, positioning, calc, shadings, shapes.geometric, shadows.blur, fadings, fit, shapes, backgrounds}

\definecolor{gainRed}{RGB}{215, 48, 39}
\definecolor{lossBlue}{RGB}{69, 117, 180}
\definecolor{deepSpace}{RGB}{40, 40, 50}
\definecolor{gridLine}{RGB}{200, 200, 220}

\usepackage{mathtools}
\usepackage{multirow}
\usepackage{subcaption}
\usepackage{url}
\usepackage[colorlinks=true, urlcolor=blue, citecolor=blue, linkcolor=blue]{hyperref}

\begin{document}
\title{Local Thermodynamics and R\'enyi Bounds on Black Hole Mergers}
\author{Sudhanva Joshi}
\email[]{sudhanvajoshi.rs.phy24@itbhu.ac.in}
\affiliation{Department of Physics, Indian Institute of Technology (Banaras Hindu University), Varanasi - 221005, India}

\author{Sunil Kumar Mishra}
\email[]{sunilkm.app@iitbhu.ac.in}
\affiliation{Department of Physics, Indian Institute of Technology (Banaras Hindu University), Varanasi - 221005, India}

\begin{abstract}
R\'enyi entropy inequalities have been proposed to strengthen the area-theorem bound on mergers of thermodynamically stable black holes in Anti de Sitter (AdS) space. The stronger bounds depend on the joint initial state, whereas masses and temperatures fix only its local thermal marginals. We quantify this distinction inside the canonical two-hole description. Correlations confined to rare energy fluctuations can leave both thermal states exactly intact and drive the trace distance and mutual information to zero, yet still change the joint R\'enyi entropy at leading semiclassical order. We determine the smallest entropy compatible with a controlled approach to statistical independence, uniformly over all positive R\'enyi orders up to one. For equal Schwarzschild--AdS$_4$ holes, the resulting optimized canonical mass constraint reduces to the area bound at leading semiclassical order. The reduction concerns only the information needed to infer a bound; it does not produce a lighter attainable remnant or a dynamical instability. Stronger merger constraints therefore require control of the initial correlations beyond single black hole thermodynamics.

\end{abstract}

\maketitle

\section{Introduction}
\label{sec:introduction}

Black hole thermodynamics constrains a merger without requiring its full spacetime evolution. Hawking's area theorem supplies a lower bound on the final horizon area, and therefore on the mass of a specified stationary remnant~\cite{hawking1971gravitational}. Identifying the horizon area with entropy connects the geometric statement to thermodynamics~\cite{bekenstein1973black,hawking1975particle}. Such constraints are especially useful when the nonlinear dynamics cannot be solved explicitly. Their force, however, depends on precisely which information about the initial holes is supplied.

The issue becomes sharper for the R\'enyi second laws proposed by Bernamonti \textit{et al.}~\cite{bernamonti2024renyi}. In the canonical description of stable Anti de Sitter (AdS) black holes, a family of entropy inequalities yields stronger final mass bounds than the ordinary area comparison. The initial entropy is obtained by adding the two one-hole R\'enyi entropies. That addition assumes an approximately factorized initial state; an assumption stated explicitly in the original work. Correlations and R\'enyi mutual information are left as open questions. A later Gauss--Bonnet--AdS application likewise adopts the additive initial entropy~\cite{kumar2026renyi}.

The independence assumption carries information that the individual masses and temperatures do not. Two holes may possess exactly the same thermal density matrices while the joint probabilities of their energy fluctuations differ. At the level of ordinary entropy, the difference is simply their mutual information. For R\'enyi orders below one, the large number of weakly occupied states amplifies the weight of small probabilities, so proximity to a product state need not justify replacing the joint R\'enyi entropy by the sum of the marginals in the semiclassical limit.

Analogous thermodynamic sensitivities are well established. R\'enyi entropies require care in questions of thermodynamic stability and continuity~\cite{lesche1982instabilities}, and small state changes can alter leading holographic R\'enyi entropies~\cite{akers2020entanglement}. In quantum thermodynamics, generalized free energies constrain allowed transformations~\cite{brandao2015second}, yet the presence of correlations can render some of those constraints redundant~\cite{lostaglio2015stochastic,muller2018correlating}. Holographic second laws further link these issues to black hole thermodynamics~\cite{bernamonti2018holographic}. These results motivate our setup without determining the joint entropy of two initially correlated thermal holes.

We therefore ask how much of the canonical merger bound follows from local thermodynamics alone when the initial state is permitted to approach the product state in a controlled fashion. We correlate the two events in which energy exceeds the cutoff, suppressing their simultaneous occurrence while preserving both thermal marginals. The local density matrices remain exactly thermal. Trace distance to the product and ordinary mutual information both vanish, and the enhancement of any measurement-outcome probability remains uniformly small. The joint R\'enyi entropy nevertheless shifts by an amount of order the black-hole entropy. Within this class, which includes arbitrary quantum states, we determine the sharp leading lower entropy and exhibit a single separable construction that attains it uniformly for all $0<n\le1$. The full family of canonical merger inequalities then yields no leading mass bound beyond the area theorem.

The statement is limited to the effective canonical two-hole description employed in the original bound. Our analysis isolates how to infer a mass constraint from incomplete preparation data without modeling the dynamical merger channel or bulk geometry. Final measurement probabilities remain close under a common quantum channel. The corresponding statement for the mean remnant mass also requires control of the final reference variance, as specified in Eq.~\eqref{eq:mean_continuity}.

Sections~\ref{sec:canonical}--\ref{sec:mass} develop the argument; Sec.~\ref{sec:interpretation} discusses its gravitational interpretation. Detailed proofs and numerical derivations appear in the appendices.

\section{Canonical entropy and the merger bound}
\label{sec:canonical}

Consider two equal, non-rotating Schwarzschild--AdS$_4$ holes of initial radius $r_i$ and AdS radius $\ell$. We set $\hbar=c=k_{\rm B}=1$ and write $x=r/\ell$, $\Lambda=\ell^2/G_N$. In the Einstein approximation, the thermodynamic quantities are
\begin{equation}
 S=\pi\Lambda x^2,\qquad
 M=\frac{\Lambda}{2\ell}x(1+x^2),\qquad
 T=\frac{1+3x^2}{4\pi\ell x}.
 \label{eq:thermo}
\end{equation}
Here, $M$ is the one-hole thermodynamic mass. The semiclassical limit is $\Lambda\to\infty$ at fixed $x_i>1$; this range selects the dominant black hole phase. Positive heat capacity already holds for $x>1/\sqrt3$, yet the black hole saddle is not globally dominant below the Hawking--Page transition~\cite{hawking1983thermodynamics}.

The one-hole state is the canonical ensemble
\begin{equation}
 \tau_\beta=\frac{e^{-\beta H}}{Z(\beta)},\qquad
 S_n(\tau_\beta)=
 \frac{\log Z(n\beta)-n\log Z(\beta)}{1-n},
 \label{eq:canonical_renyi}
\end{equation}
with the von Neumann entropy $S_1=-\Tr\tau_\beta\log\tau_\beta$ recovered by continuity. The Euclidean prescription relates the leading partition function to the thermodynamic action~\cite{hawking1983thermodynamics,gibbons1977action}. For the quantities in Eq.~\eqref{eq:thermo} one has
\begin{equation}
 \log Z(\beta)=S-\beta M
 =\pi\Lambda\frac{x^2(x^2-1)}{1+3x^2}.
 \label{eq:einstein_partition}
\end{equation}
Both Eq.~\eqref{eq:thermo} and Eq.~\eqref{eq:einstein_partition} are leading Einstein saddle expressions, not exact microscopic partition functions. The appearance of $n\beta$ shows why $n<1$ probes hotter states. Thermal partition functions likewise enter holographic R\'enyi
calculations~\cite{hung2011holographic}.

Let $F_n(M)$ denote the canonical entropy $S_n(\tau_{\beta(M)})$ along the one-hole branch. For an initial joint state $\rho_{12}$ the entropy based merger bound rests on the premise
\begin{equation}
 F_n(M_f)\ge S_n(\rho_{12}),\qquad 0<n\le1.
 \label{eq:merger_premise}
\end{equation}
We examine the consequences of this necessary condition \cite{bernamonti2024renyi}. It is not a statement that every quantum channel increases all R\'enyi entropies; the coarse-grained canonical endpoint and its entropy-monotonicity assumption remain part of the model.

Statistical independence sets $P=\tau_\beta\otimes\tau_\beta$ and yields $S_n(P)=2F_n(M_i)$. At $n=1$ the leading entropy comparison recovers the classical area bound
\begin{equation}
 x_A=\sqrt2x_i,\qquad M_A=M(x_A).
 \label{eq:area_mass}
\end{equation}
For a product state, the lower R\'enyi orders give a stronger condition.
Their optimum within the Einstein canonical family occurs at
$T_P=\sqrt2T_i$~\cite{bernamonti2024renyi}; the corresponding radius and mass are derived in Appendix~\ref{sec:figsupp}. Because the canonical state has infinite rank, the bound is understood as a limit of positive order inequalities and does not require a finite Hartley entropy $S_0$.

The comparison employs internal thermodynamic masses. The total initial energy of two separated holes in AdS also receives orbital and interaction contributions, so the bounds derived below should not be read as radiated energy fractions relative to $2M_i$.

\section{Initial correlations with unchanged local thermodynamics}
\label{sec:correlations}

The two local equilibrium states leave undetermined how the holes were prepared together. To isolate that missing information, we keep both marginals exactly equal to $\tau=\tau_\beta$. Every one-hole observable then remains unchanged, including the full energy distribution, its mean, and its fluctuations. Ordinary entropy simply records the difference as mutual
information,
\begin{equation}
 S_1(\rho_{12})=2S_1(\tau)-I(1{:}2).
 \label{eq:ordinary_defect}
\end{equation}
At other orders, the algebraic defect $2S_n(\tau)-S_n(\rho_{12})$ is not in general a divergence-based R\'enyi mutual information. The distinction becomes relevant whenever a correlation correction is inserted into a merger inequality~\cite{kudler2023renyi,kudler2024renyi}. We therefore compute the joint entropy directly.

Write $\tau=\sum_jp_j|j\rangle\langle j|$ in an energy eigenbasis and introduce an energy cutoff that partitions the spectrum into a low energy sector $L$ and its complement $H$. Let $\delta$ be the thermal probability of $H$ and set $a=1-\delta$, so that $\tau=a\tau_L+\delta\tau_H$ with both conditional states normalized. Define
\begin{equation}
 Q=(1-2\delta)\tau_L\otimes\tau_L
       +\delta(\tau_L\otimes\tau_H+\tau_H\otimes\tau_L),
 \label{eq:construction}
\end{equation}
for $0<\delta<1/2$. Both reduced states remain exactly $\tau$. The construction alters only the joint statistics of two binary events, whether each energy exceeds the cutoff. In the product state, those events are independent; in $Q$ they never occur together. No entanglement and no matching of individual microscopic labels are required. The state is diagonal in the product energy basis and is stationary under $H_1+H_2$.

The product probability of a simultaneous high-energy fluctuation is $\delta^2$. Suppressing this event and adjusting the other sector weights to preserve both marginals gives
\begin{align}
T(Q,P)&=\tfrac12\|Q-P\|_1=2\delta^2,\nonumber\\
I_Q(1{:}2)&=(1-2\delta)\log(1-2\delta)-2a\log a
\nonumber\\
&=\delta^2+O(\delta^3),\nonumber\\
\Dmax(Q\Vert P)&=-\log a.
\label{eq:closeness}
\end{align}
Here $D_{\max}(\rho\Vert P)$ is the smallest $\eta$ such that $\rho\le e^\eta P$~\cite{datta2009min}. Operationally, it means that no measurement outcome can be more than $e^\eta$ times as likely under $\rho$ as under $P$. This bound controls rare outcomes in the forward direction without bounding likelihood ratios from below.

For a stable thermal state, the cutoff can sit well above a typical energy fluctuation while remaining close to the mean on the scale of the black hole mass. Concretely we take
\begin{equation}
 \beta(E_c-\langle H\rangle)=d,\qquad
 \sqrt\Lambda\ll d\ll\Lambda.
 \label{eq:physical_cutoff}
\end{equation}
Under the concentration condition stated later, $\delta\to0$ and all three quantities in Eq.~\eqref{eq:closeness} vanish. Because the marginals are held fixed, the total expectation value of $H_1+H_2$ is likewise unchanged.

A simple preparation exists within the effective bipartite model: measure the two energy sector flags without resolving the microstate and accept with probability $aQ/P$ in each sector. The accepted state is $Q$, and the success probability approaches 1. The procedure is a coordinated, heralded filter, not a freely available thermal operation. This shows that microscopic label correlations are unnecessary, while leaving the gravitational formation mechanism unspecified. Exact preparation details and information identities are collected in Appendix~\ref{sec:state_details}.

\section{The joint entropy carried by rare fluctuations}
\label{sec:entropy}

The effect of the construction on a R\'enyi bound appears before any gravitational limit is taken. Let $L_n=\sum_{j\in L}p_j^n$ and $H_n=\sum_{j\in H}p_j^n$. Orthogonality of the energy sectors supplies the exact power sum
\begin{equation}
 \Tr Q^n=(1-2\delta)^na^{-2n}L_n^2+2a^{-n}L_nH_n.
 \label{eq:moment}
\end{equation}
The product state would instead yield $(L_n+H_n)^2$. Although $H_1=\delta$ is small, $H_n$ need not be small for $n<1$: a fractional power enhances the relative weight of the many weakly occupied states. The missing $H_n^2$ term therefore contributes at leading entropy order even when the removed ordinary probability is negligible.

To make the semiclassical statement precise, we consider a family of faithful Gibbs states $\tau_\Lambda$ with $Z_\Lambda(n\beta)<\infty$ for every $n>0$ and with $S=S_1(\tau_\Lambda)=s\Lambda+o(\Lambda)$, $s>0$. The thermal energy fluctuations are required to remain extensive in a fixed hotter neighborhood,
\begin{equation}
 \sup_{0\le u\le u_0}\beta^2
 \operatorname{Var}_{\tau_{(1-u)\beta}}(H_\Lambda)
 \le c_*\Lambda,
 \label{eq:fluctuations}
\end{equation}
where $c_*>0$ and $0<u_0<1$ are independent of $\Lambda$. This is the standard single-phase fluctuation scale; a phase-coexistence distribution need not satisfy it. Large Schwarzschild--AdS holes at fixed $x_i>1$ obey the bound inside the dominant Einstein saddle.

Let the allowed departure from independence vanish according to $\eta_\Lambda>0$, $\eta_\Lambda\to0$ and $\log(1/\eta_\Lambda)=o(\Lambda)$. Set $P=\tau_\Lambda \otimes \tau_\Lambda$. The states consistent with this information form the class
\begin{equation}
 \calC_\Lambda=\{\rho:\rho_1=\rho_2=\tau_\Lambda, \ \rho\le e^{\eta_\Lambda}P\}.
 \label{eq:class}
\end{equation}
The condition is stronger than a mere small-trace-distance requirement. Even so, the smallest attainable joint entropy is
\begin{equation}
 \inf_{\rho\in\calC_\Lambda} S_n(\rho)
 =S_1(\tau_\Lambda)+S_n(\tau_\Lambda)+o(\Lambda),
 \qquad 0<n\le1.
 \label{eq:sharp}
\end{equation}
The error term is uniform in $n$, and a single sequence of the states defined in Eq.~\eqref{eq:construction} attains the right-hand side for every index simultaneously. The lower bound holds for every quantum state in the class, including entangled states and states that fail to commute with $P$.

Two spectral properties explain this result. The normalized low-energy state satisfies $S_n(\tau_L)\le\log\operatorname{rank}\tau_L\le S+d$, so its entropy is bounded by the ordinary thermal entropy up to a subleading correction. The mixed low--high sectors retain the hotter spectral weight of one thermal factor. Together with Eq.~\eqref{eq:moment}, these observations give the upper bound. For the quantum lower bound, operator antitonicity of $x^{n-1}$ for $0<n<1$ gives
\begin{equation}
\Tr\rho^n\ge e^{-(1-n)\eta_\Lambda}\Tr(\rho P^{n-1}),
\qquad 0<n<1.
\label{eq:operator_lower}
\end{equation}
Concentration of one marginal then leaves one ordinary-entropy factor and the full R\'enyi entropy of the other. The complete proof, including control of the $n\to1$ region, is given in Appendix~\ref{app:uniform}. A convenient cutoff is $d=[2c_*\Lambda\log(1/(1-e^{-\eta_\Lambda}))]^{1/2}$; the residual error is then bounded by $d+O(\sqrt\Lambda)+O(\eta_\Lambda)$.

Exact zeros in the high-high sector are unnecessary. At countably infinite rank, one can restore strictly positive weights there while preserving both marginals and the operator bound, and still obtain the same uniform entropy \eqref{eq:sharp}. Such states are constructed in Appendix~\ref{app:fullsupport}. This distinction is essential: at a fixed finite dimension, a full support Hartley entropy is fixed by the rank.

This scaling remains fully consistent with fixed-state continuity theorems. For fixed $\Lambda$ the assumption of finite positive moments recovers the classical fixed-marginal continuity criterion~\cite{kovavcevic2015entropy}. An ordinary mean energy bound alone does not, in general, guarantee R\'enyi continuity for $0<n<1$; existing infinite-dimensional bounds retain spectral or higher-moment control~\cite{becker2023classical}. The present effect is a nonuniform thermodynamic limit: the state approaches a product while the number of relevant microstates grows. Related mathematical settings appear in the constrained-coupling and smoothing literature (see Appendix~\ref{app:uniform}); the quantity required here is the uniform thermal entropy of Eq.~\eqref{eq:sharp}.

\section{Consequence for the final-mass constraint}
\label{sec:mass}

We return to the merger. The area comparison fixes the mass $M_A$ by the condition $S(M_A)=2S(M_i)$. To test the full R\'enyi family at this mass it is useful to differentiate the canonical entropy with respect to the one-hole mass. The first law together with Eq.~\eqref{eq:canonical_renyi} gives
\begin{equation}
 \partial_M F_n
 =\frac{\beta}{C_V(T)}
   \frac{\int_1^{1/n}C_V(Tu)\,du}{n^{-1}-1}.
 \label{eq:response}
\end{equation}
The $n=1$ limit recovers $\partial_M S=\beta$. Whenever the positive heat capacity is isotonic for all $T\ge T_i$, the average that appears in Eq.~\eqref{eq:response} is at least $C_V(T)$, and therefore $\partial_M F_n\ge\beta$. Integrating from $M_i$ to $M_A$ yields
\begin{equation}
 F_n(M_A)\ge F_n(M_i)+S(M_i),\qquad 0<n\le1.
\label{eq:area_suffices}
\end{equation}
The area theorem remnant thus carries enough canonical entropy to satisfy the leading initial entropy of Eq.~\eqref{eq:sharp} at every index.

For Schwarzschild--AdS$_4$, the heat capacity is
\begin{equation}
 C_V=2\pi\Lambda x^2\frac{1+3x^2}{3x^2-1},
 \label{eq:heat_capacity}
\end{equation}
which is positive and isotonic on $x\ge1$. The required response property, therefore, holds in the Einstein canonical model. The same state attains Eq.~\eqref{eq:sharp} for all $n$, so no interchange of state minimization and index optimization is needed. In a fixed relative neighborhood of $M_A$ one has $T/M=O(\Lambda^{-1})$, and the uniform $o(\Lambda)$ entropy error can therefore be absorbed by an $o(M_A)$ mass increase.

We define the mass threshold obtained from the proposed inequalities (with the area condition imposed)
\begin{equation}
 \begin{split}
 \mathcal B[\rho]=\inf\{M\ge M_A:\\
 F_n(M)\ge S_n(\rho)\ \text{for all }0<n\le1\}.
 \end{split}
 \label{eq:threshold}
\end{equation}
The strongest lower bound that holds for every preparation in the class Eq.~\eqref{eq:class} is then
\begin{equation}
 \inf_{\rho\in\calC_\Lambda}\mathcal B[\rho]
 =M_A+o(M_A).
 \label{eq:robust_threshold}
\end{equation}
The extra mass constraint that follows from the product entropy is therefore not fixed by the two local thermal states, even when a vanishing departure from independence is imposed in the strong sense of Eq.~\eqref{eq:class}.

\begin{figure*}
 \centering
 \includegraphics[width=0.92\textwidth]{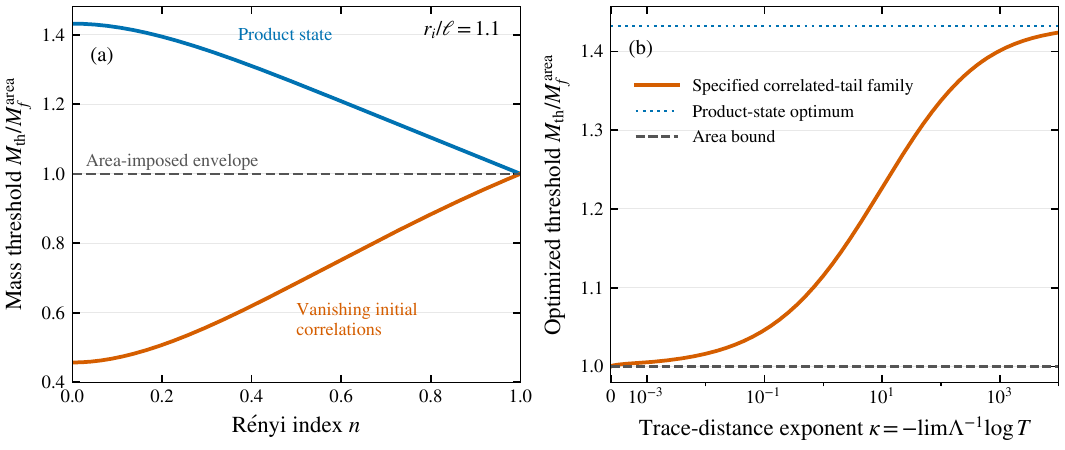}
\caption{Canonical mass constraints for two equal Schwarzschild--AdS$_4$ black holes with $r_i/\ell=1.1$. (a) The product initial state gives the blue R\'enyi thresholds. The orange thresholds use the correlated entropy $S_1(\tau)+S_n(\tau)$; after the area condition is imposed, their upper envelope is the dashed line. The $n=0$ point denotes a limit of positive orders. (b) The optimized threshold within the binary energy-sector construction as a function of $\kappa=-\lim_{\Lambda\to\infty}\Lambda^{-1}\log T(Q,P)$. The origin describes subexponential precision; the axis is logarithmic above $10^{-3}$. This curve concerns this particular family of preparations, not a sharp optimization over every correlated state. Both panels show necessary entropy-derived constraints, not attainable remnant masses.}
\label{fig:thresholds}
\end{figure*}

Figure~\ref{fig:thresholds} quantifies the distinction. At $r_i/\ell=1.1$ the product prescription yields $M_P/M_A=1.43227$. For the correlated construction, the complete entropy family, together with the area theorem, yields a leading ratio of 1. Panel (a) also displays the entropy-only thresholds before the area condition is imposed; values below unity do not constitute violations of the area theorem.

Panel (b) examines how much more restrictive the preparation information must become before an additional mass bound reappears inside the same construction. A positive $\kappa$ corresponds to an exponentially small trace distance on the entropy scale. As $\kappa$ grows, the energy cutoff moves deeper into the tail, and the optimized constraint approaches the product result. The curve is obtained directly from the Schwarzschild--AdS saddle, without any fitted spectrum or finite-size extrapolation. Its derivation and numerical validation appear in Appendix~\ref{sec:figsupp}. This curve characterizes this specific construction rather than bounding arbitrary correlated states.

The all-index gravitational conclusion requires more than local stability. Eq.~\eqref{eq:fluctuations} controls temperatures only near $T_i$, while $n\to0$ samples arbitrarily high temperatures through Eq.~\eqref{eq:canonical_renyi}. The Einstein saddle satisfies the response condition throughout that range. Extending the result to an exact microscopic theory further demands uniform control over the hotter branch and local agreement between its entropy--mass relation and the area law. A fixed-index semiclassical expansion alone does not establish these properties. Appendix~\ref{sec:uniform_thermodynamic_bridge} records sufficient conditions that go beyond positive heat capacity.

\section{What preparation information is missing?}
\label{sec:preparation}

The onesided likelihood condition in Eq.~\eqref{eq:class} allows severe suppression of small joint probabilities; a freedom directly exploited by Eq.~\eqref{eq:construction}. A lower bound of the form $\rho\ge\lambda P$, with $\lambda>0$ fixed for each state under consideration, supplies qualitatively different information: it prevents the joint spectrum from becoming arbitrarily smaller than the product spectrum and thereby restores the extreme-index product constraint of the Einstein model (Appendix~\ref{sec:state_details}). Full support alone does not furnish such a floor. The decisive question is therefore how the rare joint probabilities are controlled, not merely whether they are nonzero.

The same distinction shows why a small correlation scalar is insufficient by itself. Ordinary mutual information in Eq.~\eqref{eq:closeness} vanishes, as do the standard positive-order forward R\'enyi divergences for this commuting pair. At fixed $n<1$, however, the entropy defect remains
\begin{equation}
 2S_n(\tau)-S_n(Q)=S_n(\tau)-S_1(\tau)+o(\Lambda).
 \label{eq:renyi_defect}
\end{equation}
Whenever the one-hole difference is extensive, the joint entropy that enters the merger bound stays uncertain at leading order. Subtracting a formal R\'enyi mutual information does not recover the missing spectrum.

\section{Gravitational interpretation and physical predictions}
\label{sec:interpretation}

The relaxed mass threshold does not signal a physical instability of the remnant. Suppose the product preparation and a correlated state are sent through the same completely positive, trace-preserving channel $\calE$ and then subjected to the same final mass measurement. If the outcome distributions are $q_P$ and $q_\rho$, positivity immediately gives $q_\rho\le e^{\eta_\Lambda}q_P$. Hence
\begin{equation}
 |\langle M_f\rangle_\rho-\langle M_f\rangle_P|
 \le\sqrt{(e^{\eta_\Lambda}-1)\operatorname{Var}_P(M_f)}.
 \label{eq:mean_continuity}
\end{equation}
Whenever the reference variance is finite, the inequality holds throughout the allowed class and does not require canonical final states. Its proof is simply the Cauchy--Schwarz inequality applied to the outcome likelihood ratio. Vanishing initial $\eta_\Lambda$, therefore, cannot generate a macroscopic shift in the mean mass once the reference standard deviation is subextensive. A change in an entropy derived lower bound and a change in an actual output distribution are distinct questions.
Appendix~\ref{sec:state_details} gives the proof and the sharper bound for the binary construction $Q$.

Gravitational interpretation of the initial state is subject to further limits. Exact tensor product localization is not automatic in gravity: constraints and gravitational dressing connect regions that appear separated~\cite{donnelly2018gravitational}. We work with an effective two-hole split of the same kind already required for the product entropy, and we demand that the split remain meaningful in the spectral tails used by the construction. No smooth bulk dual of $Q$ or of its full support modification is provided. Recent examples demonstrate that distinct quantum states can share the same classical spacetime limit and the same leading von Neumann entropy while differing in their R\'enyi entropies. Those examples underscore the distinction between geometric data and the full state~\cite{bousso2026quantum}; they do not establish the gravitational admissibility of the particular preparations considered here.

The analysis concerns correlations already present on the slice at which the entropy bound is applied. Correlations generated later during infall do not invalidate a constraint that was correctly imposed on an earlier product state. Conversely, a claim based solely on local thermodynamic data on a later slice cannot tacitly import independence from an unspecified preparation history.

Finally, the canonical entropy argument must be kept separate from the geometric Hawking--R\'enyi construction of Ref.~\cite{bernamonti2024renyi}. Replica geometries can encode information absent from the individual one-hole partition functions; geometric treatments of R\'enyi quantities make the distinction explicit~\cite{hung2011holographic,dong2016gravity}. We have not constructed the joint correlated replica saddle and therefore cannot rule out an independent constraint that might arise from it. Our result concerns the preparation information required by the canonical mass bound. It does not modify the classical area theorem or the generalized second law for semiclassical horizons~\cite{wald2001thermodynamics,wall2012proof}.

\section{Discussion}
\label{sec:discussion}

Local equilibrium fixes the entropy of each hole, yet leaves freedom in the joint statistics of their rare energy fluctuations. The construction above makes that freedom quantitative: the marginal states remain exact, ordinary correlations vanish, and measurement probabilities obey a uniform forward bound, while the joint R\'enyi entropy can still fall short of the product value by a leading semiclassical amount. Eq.~\eqref{eq:sharp} fixes the smallest attainable entropy, and a single preparation reaches it uniformly for every positive order up to one. Within the Einstein Schwarzschild--AdS$_4$ canonical model, this is already sufficient to eliminate the extra mass constraint beyond the area theorem. The stronger leading lass bound therefore cannot be inferred from the one-hole thermal states and a subexponentially vanishing tolerance $\eta_\Lambda$ in Eq.~\eqref{eq:class} alone.

Stronger conclusions demand further physical input---a preparation principle that controls the joint tails, a justified lower spectral bound, or an explicit calculation of the correlated gravitational replica state. Smoothed entropy prescriptions raise a related but distinct question, each with its own operational accuracy and limiting procedure \cite{anshu2020partially,bao2026capacity}. Which of these conditions is selected by an actual black-hole formation process is a dynamical problem. The canonical entropy calculation isolates precisely where that information is required.

\section*{Data Availability}
Data that supports findings in this paper are openly available at~\cite{joshi2025witp}

\appendix
\section{Uniform quantum lower bound and attainment}
\label{app:uniform}

The argument works with the precise state class of Eq.~\eqref{eq:class}. Fixed marginal entropy optimization has classical antecedents \cite{yu2018asymptotic,ma2025efficient}. For diagonal states, the upper bound is a capacity-constrained coupling condition whose geometry was examined in~\cite{korman2013insights}. Partially smoothed information measures likewise impose marginal restrictions~\cite{anshu2020partially}. What is required here is the sharp thermal value itself, its lower bound for noncommuting states, and a single construction that attains it uniformly throughout $0<n\le1$.

Write $Y_j=-\log p_j$ and define $K(u)=\log\Tr\tau^{1-u}-uS$. Eq.~\eqref{eq:fluctuations} implies $K(0)=K'(0)=0$ and $K(u)\le c_*\Lambda u^2/2$. In particular $\operatorname{Var}(Y)\le c_*\Lambda$ and
\begin{equation}
 S_{1-u}(\tau)\le S+\tfrac12c_*\Lambda u.
 \label{eq:end_cumulant}
\end{equation}
All joint moments remain finite: the bound $\rho\le e^\eta P$ yields $\Tr\rho^n\le e^{n\eta}(\Tr\tau^n)^2$ by eigenvalue monotonicity. Set $t=1-n$ and let $\Pi$ project onto the set $-\log p_j\ge S-h$ with $h=2\sqrt{c_*\Lambda}$. Chebyshev's inequality gives $\zeta=\Tr[\tau(1-\Pi)]\le1/4$.

To justify Eq.~\eqref{eq:operator_lower} when $\rho$ may have a kernel, apply operator antitonicity to $\rho+\epsilon\mathbf{1}\le e^\eta(P+\epsilon\mathbf{1})$, multiply by $\rho$ and take the trace. Monotone convergence as $\epsilon\downarrow0$ recovers the inequality without ever tracing $\epsilon\mathbf{1}$. Next, $\tau^{-t}\Pi\ge e^{t(S-h)}\Pi$ and
\begin{align}
 \Tr[\rho(\tau^{-t}\otimes\Pi)]
 &=\Tr\tau^n-\Tr[\rho(\tau^{-t}\otimes(1-\Pi))]
 \nonumber\\
 &\ge(1-e^\eta\zeta)\Tr\tau^n.
\end{align}
The exact first marginal supplies the equality; the operator upper bound on $\rho$ supplies the inequality. Consequently
\begin{equation}
 \Tr\rho^n\ge e^{t(S-h-\eta)}(1-e^\eta\zeta)\Tr\tau^n.
 \label{eq:universal_lower}
\end{equation}
For large enough $\Lambda$, one has $1-e^\eta\zeta\ge1/2$.

When $1-n\ge\Lambda^{-1/2}$ this yields
\begin{equation}
 S_n(\rho)\ge S+S_n(\tau)-\eta -(2\sqrt{c_*}+\log2)\sqrt\Lambda.
\end{equation}
For smaller $1-n$ combine Eq.~\eqref{eq:end_cumulant} with $S_n(\rho)\ge S_1(\rho)=2S-D(\rho\Vert P)\ge2S-\eta$. The uniform lower error is therefore at most $\eta+C_-\sqrt\Lambda$, where $C_-=\max\{2\sqrt{c_*}+\log2,\ c_*/2\}$. The argument nowhere assumes that $\rho$ commutes with $P$.

For attainment at tolerance $\eta=\eta_\Lambda$ choose
\begin{equation}
 d=\sqrt{2c_*\Lambda\log\frac1{1-e^{-\eta}}}.
 \label{eq:end_cutoff}
\end{equation}
Then $d=o(\Lambda)$, $d/\sqrt\Lambda\to\infty$, and the Chernoff bound at $u=d/(c_*\Lambda)<u_0$ gives $\delta\le1-e^{-\eta}$. If the tail is empty, set $Q=P$; the same rank bound then yields $S_n(\tau)-S\le d$. Otherwise $\delta>0$. Every retained probability is at least $e^{-S-d}$. Thus $L=\{j:-\log p_j\le S+d\}$ has rank $r_L\le e^{S+d}$. Writing $P_n=\Tr\tau^n$, for $0<n<1$ we have
\begin{equation}
a^{-n}L_n=\Tr\tau_L^n\le r_L^{1-n},
\end{equation}
and Eq.~\eqref{eq:moment}
gives
\begin{equation}
\Tr Q^n
\le 2a^{-n}L_nP_n
\le 2e^{(1-n)(S+d)}P_n.
\end{equation}For $1-n\ge\Lambda^{-1/2}$, the upper error is at most $d+\log2\sqrt\Lambda$. Inside the boundary layer, $S_n(Q)\le S_{1-\Lambda^{-1/2}}(Q)$ together with Eq.~\eqref{eq:end_cumulant} produces
\begin{equation}
 S_n(Q)\le S+S_n(\tau)+d+(\log2+c_*/2)\sqrt\Lambda.
 \label{eq:end_upper}
\end{equation}
At $n=1$, ordinary subadditivity supplies the upper bound directly. The construction, therefore, attains the bound simultaneously for all indices, without any interchange of state minimization and index optimization.

\section{Full support with the same sharp envelope}
\label{app:fullsupport}

Assume countably infinite rank and order the probabilities $p_j$ decreasingly. Choose a symmetric, strictly positive matrix $W$ on $H\times H$ that satisfies
\begin{equation}
 W_{ij}\le p_ip_j,\quad
 \sum_{i,j\in H}W_{ij}^n\le P_n:=\sum_jp_j^n
 \quad(n>0).
 \label{eq:packing}
\end{equation}
Such a matrix exists for all $n$ simultaneously. Enumerate the unordered pairs in $H\times H$ and assign to each pair a disjoint interval of thermal indices lying beyond all earlier intervals. The interval length is one for a diagonal pair and two otherwise. Place the interval far enough that its last probability satisfies $p_k\le p_ip_j$, and set $W_{ij}=W_{ji}=p_k$. The one or two copies of $p_k^n$ are then bounded by the power sum over the assigned interval. Summing the disjoint intervals establishes Eq.~\eqref{eq:packing}. The construction uses only the facts that $p_k>0$ and $p_k\to0$.

Let $w_i=\sum_{j\in H}W_{ij}$ and $w=\sum_iw_i\le\delta^2$. Define
\begin{equation}
 q_{ij}=\begin{cases}
 (1-2\delta+w)p_ip_j/a^2,&i,j\in L,\\
 (p_i/a)(p_j-w_j),&i\in L,\ j\in H,\\
 (p_j/a)(p_i-w_i),&i\in H,\ j\in L,\\
 W_{ij},&i,j\in H.
 \end{cases}
 \label{eq:packed_state}
\end{equation}
Because $w_i\le\delta p_i$, every entry is positive. The row and column sums are exactly $p$. The likelihood ratios are at most one in the $LL$ and $HH$ sectors and lie in the interval $[1,1/a]$ in the cross sectors. It follows that
\begin{equation}
 \Dmax(q\Vert P)\le-\log a,\qquad
 T(q,P)=2(\delta^2-w)\le2\delta^2.
\end{equation}
The first three sectors contribute at most $2a^{-n}L_nP_n$ to the $n$th moment, while the $HH$ sector contributes at most $P_n$. For $0<n<1$, the same rank bound therefore gives
\begin{equation}
 \Tr q^n\le3e^{(1-n)(S+d)}P_n.
\end{equation}
Eq.~\eqref{eq:end_upper} continues to hold with $\log2$ replaced by $\log3$. Combined with the universal lower bound, this yields the same sharp envelope, even when the admissible class is restricted to stationary, separable, full support states.

When energy degeneracies are present, enumerate pairs of entire eigenspaces and assign intervals of lengths $g_E^2$ or $2g_Eg_{E'}$. Assign the same diagonal probability to every microstate pair within each product energy block. The corresponding density-operator block is therefore proportional to $\Pi_{E}\otimes\Pi_{E^{'}}$. Finite Gibbs degeneracies keep the construction well-defined and invariant under basis changes within each eigenspace. The resulting filter depends on both energies and requires detailed spectral information; it is not the two-flag protocol in the main text. At a fixed finite dimension, a full support Hartley entropy is fixed by the rank, so the infinite-rank conclusion cannot be imported unchanged.

\section{Exact state identities and spectral-tail conditions}
\label{sec:state_details}

We adopt the notation of the main text. Set $P=\tau\otimes\tau$, $a=1-\delta$, and condition $\tau$ on the two orthogonal energy sectors. Unless otherwise stated, $0<\delta<1/2$. The binary construction assigns sector probabilities $(1-2\delta,\delta,\delta,0)$ in the order $LL,LH,HL,HH$, while the product state has $(a^2,a\delta,a\delta,\delta^2)$. The conditional states coincide on every occupied sector. The reduced state of either subsystem is $(1-2\delta+\delta)\tau_L+\delta\tau_H=\tau$.

The likelihood ratio $R=Q/P$ reads
\begin{equation}
 R_{LL}=\frac{1-2\delta}{a^2},\qquad
 R_{LH}=R_{HL}=\frac1a,\qquad R_{HH}=0.
 \label{eq:supp_likelihood}
\end{equation}
Consequently, the trace distance, mutual information, maximum relative entropy, and Pearson divergence are given exactly by
\begin{align}
 T(Q,P)&=2\delta^2,\\
 I&=(1-2\delta)\log(1-2\delta)-2a\log a\nonumber\\
  &=\delta^2+O(\delta^3),\\
 D_{\max}(Q\Vert P)&=-\log a,\\
 \chi^2(Q\Vert P)&=\sum_{ij}P_{ij}(R_{ij}-1)^2
                 =\frac{\delta^2}{a^2},
 \label{eq:supp_binary_identities}
\end{align}
where $I=D(Q\Vert P)$. The entropy identity $I=2S(\tau)-S(Q)$ follows at once from the exact marginals. For this commuting pair, every standard positive order forward R\'enyi divergence is bounded by $D_{\max}$. (The same bound need not hold for all quantum Petz orders when the states fail to commute.)

The heralded preparation is a diagonal filter that accepts each input energy pair with probability $aR$. The map is valid because $0\le aR\le1$, succeeds with probability $a$, and yields $Q$ upon success. The failure state $(P-aQ)/\delta$ likewise has both marginals exactly equal to $\tau$. The filter preserves energy eigenstates but requires classical coordination and post-selection; it is not claimed to be a freely available operation inside a gravitational thermal resource theory.

For the full-support state constructed in Appendix~\ref{app:fullsupport}, the analogous acceptance rule $ a q_{ij}/(p_ip_j)$ prepares the state with probability $a$. Its implementation, however, demands a detailed joint spectral filter.

\subsection{An additional trace-distance tolerance}

An independent trace-distance tolerance $\epsilon_\Lambda\to0$ may be imposed by choosing $d^2=2c_*\Lambda$ times
\begin{equation}
 \max\left\{\log\frac1{1-e^{-\eta_\Lambda}},   \tfrac12\log\frac2{\epsilon_\Lambda}\right\}.
\end{equation}
Whenever both tolerances decay subexponentially, the upper error remains $o(\Lambda)$ while $T\le\epsilon_\Lambda$ and $\Dmax\le\eta_\Lambda$.

\subsection{The other endpoint of the entropy interval}

Every state in the class obeys
\begin{equation}
 S_n(\rho)\le2S_n(\tau)+\frac{n\eta}{1-n}
 \qquad(0<n<1)
 \label{eq:supp_upper_raw}
\end{equation}
by eigenvalue monotonicity. Let $v=\sqrt{\eta/(c_*\Lambda)}<u_0$. When $1-n\ge v$, the final term is at most $\sqrt{c_*\Lambda\eta}$. When $1-n<v$, entropy-order monotonicity, the bound at $1-v$ and the estimate $S_{1-v}(\tau)\le S+c_*\Lambda v/2$ together yield
\begin{equation}
 S_n(\rho)\le2S_n(\tau)+2\sqrt{c_*\Lambda\eta},
 \qquad 0<n\le1.
 \label{eq:supp_sharp_upper}
\end{equation}
At $n=1$, ordinary subadditivity is enough. The product state belongs to the class and attains $2S_n(\tau)$ exactly. Combined with the sharp lower theorem of the main text, both leading endpoints are therefore attained. This does not assert that every spectrum lying between the two envelopes is realized by a single interpolation. The extensive separation of the endpoints concerns fixed $0<n<1$ for which $S_n(\tau)-S(\tau)=\Theta(\Lambda)$.

\subsection{A positive product floor restores the extreme-index bound}

If $\rho\ge\lambda P$ for some $\lambda>0$, eigenvalue monotonicity supplies
\begin{equation}
 \Tr\rho^n\ge\lambda^n(\Tr\tau^n)^2,\qquad
 S_n(\rho)\ge2S_n(\tau)+\frac n{1-n}\log\lambda.
 \label{eq:supp_floor}
\end{equation}
The coefficient $\lambda$ may depend on $\Lambda$ but remains strictly positive at each fixed $\Lambda$. For the Einstein canonical family $S_n(\tau)$ diverges as $n^{-2}$ when $n\downarrow0$. Any such floor, therefore, retains the product's high-temperature entropy coefficient. The order of limits is essential: the optimization includes arbitrarily small positive $n$ at each $\Lambda$. No finite canonical Hartley entropy is assigned.

For the convex combination $Q_\lambda=(1-\lambda)Q+\lambda P$ with $0<\lambda\le1$, the product threshold is also an upper bound. To see this, define the diagonal signed measure $d=\tau_H-\tau_L$. Then
\begin{equation}
 Q=P-\delta^2 d\otimes d,
 \qquad Q_\lambda=P-t d\otimes d,
 \quad t=(1-\lambda)\delta^2.
\end{equation}
For $0<n<1$ the function $f(t)=\sum_{ij}(p_ip_j-td_id_j)^n$ is concave on the interval $0\le t\le\delta^2$ and
\begin{equation}
 f'(0)=-n\left(\sum_i p_i^{n-1}d_i\right)^2\le0.
\end{equation}
The sum is finite because it equals $H_n/\delta-L_n/a$. Concavity therefore implies $\Tr Q_\lambda^n\le(\Tr\tau^n)^2$. At $n=1$, subadditivity is used. The two bounds together establish that the optimized threshold equals the product value within the stated canonical model for every $\lambda>0$.

A product floor is sufficient but not necessary. As a counter-example, choose $p_k\le p_l$ and subtract $p_k^2$ from the diagonal entries $(k,k)$ and $(l,l)$ of $P$, adding the same amount to the off-diagonal entries $(k,l)$ and $(l,k)$. Positivity and both marginals are preserved, yet $q_{kk}=0$ forces $D_{\max}(P\Vert q)=\infty$. Only four moment terms are altered, so
\begin{equation}
 \Tr q^n=(\Tr\tau^n)^2+O(1),\qquad n\downarrow0.
\end{equation}
At infinite rank $\Tr\tau^n\to\infty$, the product coefficient survives. Moreover $D_{\max}(q\Vert P)=\log(1+p_k/p_l)$ can be made arbitrarily small. The decisive information is therefore the growth of the joint power sums, not support alone or a finite reverse maximum divergence.

By contrast, the moment upper bounds for the binary and packed-tail states are at most $C_\Lambda\Tr\tau^n$ near zero, with $C_\Lambda<\infty$ at each fixed $\Lambda$. Their universal lower bound is another positive constant times $\Tr\tau^n$. Hence
\begin{equation}
 \lim_{n\downarrow0}\frac{S_n(Q)}{S_n(\tau)}=1
\end{equation}
for either construction. In finite dimension $d$, a faithful joint state instead has $S_0=2\log d$; its difference from $S(\tau)+S_0(\tau)$ can be extensive. This is the reason for the infinite-rank qualification of the full support uniform theorem.

\subsection{Observable stability and discrimination}

The main text stability inequality holds for the entire quantum class, not merely the binary state. For any fixed completely positive trace preserving channel $\mathcal E$, operator monotonicity yields $\mathcal E(\rho)\le e^\eta\mathcal E(P)$. Any subsequent mass measurement, therefore, produces probability measures satisfying $q\le e^\eta p$. Writing $R=dq/dp$ one has $0\le R\le e^\eta$, $\mathbb E_p R=1$ and
\begin{equation}
 \chi^2(q\Vert p)=\mathbb E_p R^2-1\le e^\eta-1.
\end{equation}
For an arbitrary observable $A$ of finite reference variance,
\begin{align}
 |\mathbb E_q A-\mathbb E_p A|
 &=|\mathbb E_p[(R-1)(A-\mathbb E_p A)]|\nonumber\\
 &\le\sqrt{(e^\eta-1)\operatorname{Var}_p(A)}.
\end{align}
Neither commutativity of the input nor a canonical form of the output is required. Finite reference variance also guarantees that the necessary moments remain finite under $q$. The comparison is performed with unconditional outputs. If the channel is post-selected and each output is renormalized separately, a success probability factor must be restored; the bound displayed above cannot then be used unchanged.

The binary construction admits the sharper estimates that follow. A fixed quantum channel followed by a measurement acts as a classical Markov kernel on the diagonal input labels. The likelihood bound of Eq.~\eqref{eq:supp_likelihood} therefore implies, for every final event $B$,
\begin{equation}
 \Pr_Q(B)\le\frac{\Pr_P(B)}{a}.
\end{equation}
Classical $\chi^2$ data-processing and the Cauchy--Schwarz inequality then give
\begin{equation}
 |\mathbb E_Q M_f-\mathbb E_P M_f|
 \le\frac{\delta}{a}\sqrt{\operatorname{Var}_P(M_f)}
\end{equation}
whenever the reference variance is finite. These statements apply to the binary state independently of any canonical-output assumption. The packed full support state retains the likelihood bound, although the exact $\chi^2$ formula of the binary construction is not claimed for that extension.

The binary preparation also possesses an exact discrimination scale. Tensor product telescoping yields $T(P^{\otimes N},Q^{\otimes N})\le2N\delta^2$, so a constant advantage already requires $N=\Omega(\delta^{-2})$ even for optimal joint measurements. Conversely, the observation of an $HH$ event certifies the product: its probability under $P^{\otimes N}$ is $1-(1-\delta^2)^N$ and vanishes under $Q^{\otimes N}$. Hence $N=O(\delta^{-2})$ suffices for any fixed error strictly less than one half, and the optimal scaling is $N=\Theta(\delta^{-2})$. For a fixed extensive cutoff, one has $\delta=\exp[-\Lambda J+o(\Lambda)]$, so the number of copies carries exponent $2J$. The statement concerns the discrimination of preparations; it is not offered as a new general lower bound on entropy estimation.

\section{Conditions for an exact canonical extension}
\label{sec:uniform_thermodynamic_bridge}

The Einstein calculation uses the canonical response over arbitrarily high temperatures. For an exact canonical extension, assume the hypotheses of Appendix~\ref{app:uniform}, differentiable Gibbs thermodynamics, finite partition functions for every $T\ge T_i$, and a positive heat capacity $C_\Lambda(T)$ that is nondecreasing throughout this range. Writing $F_{n,\Lambda}(T)=S_n(\tau_{\Lambda,T})$, the identity
\begin{equation}
F_{n,\Lambda}(T)
=\frac{1}{n^{-1}-1}
\int_1^{1/n}S_\Lambda(Tu)\,du
\end{equation}
gives
\begin{equation}
\frac{\partial F_{n,\Lambda}}{\partial E_\Lambda}
=\frac{1}{T C_\Lambda(T)}
\frac{\int_1^{1/n}C_\Lambda(Tu)\,du}{n^{-1}-1}
\ge\frac1T .
\end{equation}
These expressions hold for $0<n<1$; the $n=1$ response is the first law.

Let $M_{2,\Lambda}$ denote the entropy-doubling mass, $S_\Lambda(M_{2,\Lambda})=2S_\Lambda(M_{i,\Lambda})$. Integration yields
\begin{equation}
F_{n,\Lambda}(M_{2,\Lambda})
\ge F_{n,\Lambda}(M_{i,\Lambda}) +S_\Lambda(M_{i,\Lambda}), \qquad 0<n\le1.
\end{equation}
The attaining joint state exceeds the right-hand side by at most $w_\Lambda=o(\Lambda)$, uniformly in $n$. Choose a fixed relative interval above $M_{2,\Lambda}$ on which $T(M)/M=O(\Lambda^{-1})$ uniformly, and let $T_+$ be its maximum temperature. Then $\Delta M_\Lambda=T_+w_\Lambda=o(M_{2,\Lambda})$ eventually lies inside that interval and supplies the required entropy for every index:
\begin{equation}
F_{n,\Lambda}(M_{2,\Lambda}+\Delta M_\Lambda) -F_{n,\Lambda}(M_{2,\Lambda}) \ge\frac{\Delta M_\Lambda}{T_+}=w_\Lambda .
\end{equation}

To identify this mass with geometry, additionally assume local uniform convergence of the rescaled entropy and mass functions to their Einstein limits on a fixed temperature interval containing the initial and area temperatures and a neighborhood above the latter. These conditions provide the required local mass regularity and imply $M_{2,\Lambda}=M_A+o(M_A)$. The canonical inequalities are therefore satisfied at $M_A+o(M_A)$ when the geometric area condition is imposed independently.

The nondecrease of the exact heat capacity across all higher temperatures remains an additional physical assumption. It follows neither from positive heat capacity alone nor from a fixed-temperature semiclassical expansion. The product benchmark in Fig.~\ref{fig:thresholds} is an Einstein-model calculation, as detailed in Appendix~\ref{sec:figsupp}.

\section{Deterministic calculation of the figure}
\label{sec:figsupp}

The figure evaluates the Einstein canonical Schwarzschild--AdS$_4$ thermodynamics together with the explicitly specified binary-sector construction. All plotted quantities are leading semiclassical entropy-derived mass thresholds. No microscopic spectrum is fitted, no finite-size extrapolation is performed, and no threshold is claimed to be dynamically attainable. Throughout the section $x_i=1.1$ unless otherwise stated. This value lies strictly inside the dominant large-black-hole phase and avoids employing a phase-coexistence saddle as the initial canonical state.

\subsection{Canonical functions and panel (a)}

Introduce the dimensionless functions
\begin{align}
 s(x)&=\pi x^2,\nonumber\\
 e(x)&=\frac{x(1+x^2)}{2},\nonumber\\
 b(x)&=\frac{4\pi x}{1+3x^2},\nonumber\\
 z(x)&=s(x)-b(x)e(x)
       =\frac{\pi x^2(x^2-1)}{1+3x^2}.
 \label{eq:figsupp_thermo}
\end{align}
Thus $S_{\rm BH}=\Lambda s$, $M=(\Lambda/\ell)e$, $\beta=\ell b$ and $\log Z=\Lambda z$ at the stated saddle order. The large-hole inverse-temperature map is
\begin{equation}
 \mathcal X(b)=
 \frac{2\pi+\sqrt{4\pi^2-3b^2}}{3b}.
 \label{eq:figsupp_inverse_temperature}
\end{equation}
Writing $y=\mathcal X(nb(x))$, the canonical R\'enyi entropy per $\Lambda$ may be evaluated as
\begin{align}
 f_n(x)
 &=\frac{z(y)-nz(x)}{1-n}\nonumber\\
 &=\frac{\pi xy(x^2+xy+y^2-1)}{3xy-1}.
 \label{eq:figsupp_stable_entropy}
\end{align}
The second form follows algebraically from $b(y)=nb(x)$ and remains regular at $n=1$, where it reduces to $f_1(x)=s(x)$. It avoids cancellation between partition functions near that endpoint. These are precisely the canonical thermodynamic functions employed for the product merger bounds in~\cite{bernamonti2024renyi}.

For each $0<n\leq1$ panel (a) solves the monotone root equations
\begin{align}
 f_n(x_f^{P})&=2f_n(x_i),\nonumber\\
 f_n(x_f^{Q})&=s(x_i)+f_n(x_i).
 \label{eq:figsupp_panel_a_roots}
\end{align}
The vertical values are $e(x_f)/e(x_A)$ with $x_A=\sqrt2x_i$. The area imposed envelope is the maximum of the entropy-derived threshold and unity. For the correlated curve, this envelope equals unity at leading order. The orange curve follows the sharp $S_1+S_n$ entropy envelope proved in the main text.

The plotted $n=0$ entries are positive-index limits, not finite canonical Hartley entropies. They follow from the high-temperature expansion $f_n(x)\sim16\pi^3/[27n^2b(x)^2]$:
\begin{equation}
 x_f^P(0^+)=\mathcal X\!\left(\frac{b_i}{\sqrt2}\right),
 \qquad x_f^Q(0^+)=x_i.
 \label{eq:figsupp_index_zero_limits}
\end{equation}
The product supremum occurs at this endpoint. A direct check is provided by the thermal average identity
\begin{equation}
 S_n(T_H)=\frac{1}{n^{-1}-1}
          \int_1^{1/n}S_1(T_Hu)\,du.
 \label{eq:figsupp_thermal_average}
\end{equation}
Because $S_1(T_H)/T_H^2$ increases on the branch under consideration, $T_{H,f}=\sqrt2T_{H,i}$ satisfies the entire product family, while Eq.~\eqref{eq:figsupp_index_zero_limits} excludes any smaller temperature. At $x_i=1.1$ one obtains
\begin{align}
 \frac{M_P}{M_A}&=1.4322667990631492,\nonumber\\
 \frac{M_i}{M_A}&=0.4569315749772717.
 \label{eq:figsupp_benchmarks}
\end{align}

\subsection{Large-deviation family in panel (b)}

Place a fixed extensive cutoff at $E_c=(\Lambda/\ell)e(x_c)$ with $x_c>x_i$. Within the same Einstein density-of-states approximation, the canonical tail probability takes the form
\begin{align}
 \delta&=\exp[-\Lambda J(x_c)+o(\Lambda)],\nonumber\\
 J(x_c)&=b_i[e(x_c)-e(x_i)]-[s(x_c)-s(x_i)].
 \label{eq:figsupp_rate}
\end{align}
The exact trace distance of the binary construction is $T(Q,P)=2\delta^2$. Its leading trace-distance exponent is therefore
\begin{equation}
 \kappa=-\lim_{\Lambda\to\infty}\Lambda^{-1}\log T(Q,P)
        =2J(x_c).
 \label{eq:figsupp_precision_exponent}
\end{equation}
For stable numerical inversion near $x_c=x_i$ set $h=x_c-x_i$ and employ
\begin{equation}
 J(x_c)=h^2\left[
 \frac{\pi(3x_i^2-1)}{1+3x_i^2}+\frac{b_i h}{2}\right].
 \label{eq:figsupp_rate_polynomial}
\end{equation}

Let $L_n=\sum_{E_k\leq E_c}p_k^n$ and $\ell_n=\lim_{\Lambda\to\infty}\Lambda^{-1}\log L_n$. The unconstrained tilted saddle has radius $\mathcal X(nb_i)$. It reaches the cutoff at $n_c=b(x_c)/b_i$. Maximizing the canonical exponent below the cutoff yields
\begin{equation}
 \ell_n=
 \begin{cases}
 (1-n)f_n(x_i),& n\geq n_c,\\
 (1-n)s(x_c)-nJ(x_c),&0<n<n_c.
 \end{cases}
 \label{eq:figsupp_low_moment_rate}
\end{equation}
The second line is the boundary value of $s(x)-nb_i e(x)-nz(x_i)$ evaluated at $x=x_c$.

For the state $Q=(1-2\delta)\tau_L\otimes\tau_L +\delta(\tau_L\otimes\tau_H+\tau_H\otimes\tau_L)$ the exact moment expression produces the leading joint entropy
\begin{equation}
 g_n(x_c)=\lim_{\Lambda\to\infty}\frac{S_n(Q)}{\Lambda}
         =f_n(x_i)+\frac{\ell_n}{1-n}.
 \label{eq:figsupp_joint_entropy_rate}
\end{equation}
At $n=1$, one sets $g_1=2s(x_i)$ by continuity at leading order. The fixed-cutoff precision curve is
\begin{equation}
R(\kappa)=\max\!\left\{
1,\ \sup_{0<n\le1}
\frac{e(x_f(n,x_c))}{e(x_A)}
\right\}.
\label{eq:figsupp_precision_curve}
\end{equation}
where $f_n(x_f)=g_n(x_c)$ and $x_c$ is obtained from Eq.~\eqref{eq:figsupp_precision_exponent}. The origin corresponds to a moderate deviation cutoff with vanishing tail probability and $x_c\to x_i$, giving $g_n=s(x_i)+f_n(x_i)$. It does not imply that a cutoff placed exactly at the canonical mean possesses an exponentially rare tail. The figure employs a linear axis below $\kappa=10^{-3}$ and a logarithmic axis above it.

Equation~\eqref{eq:figsupp_precision_curve} is a construction-specific curve. It is not a sharp minimax relation over all couplings at fixed $\kappa$. Nor does the saddle calculation assert a uniform finite-$\Lambda$ expansion for every omitted microscopic correction, or a no-go theorem for high confidence predictions that employ smoothed quantities.

\subsection{Unique maximizer and numerical validation}

The maximum appearing in Eq.~\eqref{eq:figsupp_precision_curve} can be certified analytically rather than inferred from a grid. Define $\zeta(b)=z(\mathcal X(b))$, $b_f=b(x_f)$ and $y_j=\mathcal X(nb_j)$. On the interval $0<n<n_c$ the root equation is $F(n,x_f)=0$, where
\begin{align}
 F(n,x_f)={}&\zeta(nb_f)-nz(x_f)-\zeta(nb_i)-s(x_c)
 \nonumber\\
 &+n[b_i e(x_c)+2z(x_i)].
 \label{eq:figsupp_root_function}
\end{align}
Because $F_{x_f}>0$, the derivative of the radius threshold has the sign of
\begin{align}
 D&=s(y_f)-s(y_i)-s(x_c),\nonumber\\
 \frac{dx_f}{dn}&=\frac{D}{nF_{x_f}}.
 \label{eq:figsupp_stationarity}
\end{align}
The identity follows from $F-nF_n=D$ evaluated at a root. At any stationary point one has $dx_f/dn=0$ and
\begin{equation}
 n\frac{dD}{dn}=c(y_i)-c(y_f)<0,
 \quad c(x)=\frac{2\pi x^2(1+3x^2)}{3x^2-1}.
 \label{eq:figsupp_downward_crossing}
\end{equation}
Indeed $y_f>y_i\geq x_c>1$ and
$c'(x)=4\pi x(9x^4-6x^2-1)/(3x^2-1)^2>0$ on that range.

The endpoint signs establish existence. The high-temperature expansion $\zeta(b)=16\pi^3/(27b^2)-2\pi/3+O(b^2)$ shows that $D\to2s(x_c)>0$ as $n\to0^+$. At $n=n_c$, the threshold coincides with the product threshold. The ratio $S_1(T_{H,f}u)/S_1(T_{H,i}u)$ decreases with $u$ because $c/s=2+4/(3x^2-1)$ decreases with temperature. Its weighted average in Eq.~\eqref{eq:figsupp_thermal_average} equals 2 at the product root, so the upper endpoint value is strictly smaller than 2. Hence $D(n_c)<0$. Every zero of $D$ is a strict downward crossing by Eq.~\eqref{eq:figsupp_downward_crossing}; exactly one zero therefore exists. The same endpoint-ratio argument shows that the product branch $n\geq n_c$ is decreasing, so the unique zero is the global maximum over the complete positive-index family.

The numerical implementation first brackets each monotone mass root and then brackets the unique solution of $D=0$. Both steps employ Brent's method. Independent dense index scans supply a separate computational check. Representative values at $x_i=1.1$ are
\begin{center}
\begin{tabular}{ccc}
\hline
$\kappa$ & $n_*$ & $R(\kappa)$ \\
\hline
$0.1$ & $0.899824646552$ & $1.046306292857$ \\
$1$   & $0.762437175927$ & $1.115215756195$ \\
$10$  & $0.553288151175$ & $1.226905576913$ \\
\hline
\end{tabular}
\end{center}
The maximum scaled entropy-root residual is $4.0\times10^{-14}$. The cancellation-free formula agrees with direct partition-function evaluation to $1.6\times10^{-14}$, and the relative stationary-condition residual lies below $10^{-13}$ in the reported checks. As a separate limiting benchmark, the ordered limit $x_i\downarrow1$ reproduces $x_P=1.6881650340819934$ and $M_P/M_A=1.5318922906567805$. This check does not identify a coexistence ensemble at $x_i=1$ with a pure black hole phase.

\bibliographystyle{apsrev4-2}

\bibliography{MBL}

\end{document}